# The Price of Political Uncertainty: Evidence from the 2016 U.S. Presidential Election and the U.S. Stock Markets

Jamal BOUOIYOUR [†] and Refk SELMI [‡]

**Abstract:** There is bountiful evidence that political uncertainty stemming from presidential elections or doubt about the direction of future policy make financial markets significantly volatile, especially in proximity to close elections or elections that may prompt radical policy changes. Although several studies have examined the association between presidential elections and stock returns, very little attention has been given to the impacts of elections and election induced uncertainty on stock markets. This paper explores, at sectoral level, the uncertain information hypothesis (UIH) as a means of explaining the reaction of markets to the arrival of unanticipated information. This hypothesis postulates that political uncertainty is greater prior to the elections (relative to pre-election period) but is resolved once the outcome of the elections is determined (relative to post-election period). To this end, we adopt an event-study methodology that examines abnormal return behavior around the election date. We show that collapsing stock returns around the election result is reversed by positive abnormal return on the next day, except some cases where we note negative responses following the vote count. Although Trump's win plunges US into uncertain future, positive reactions of abnormal return are found. Therefore, our results do not support the UIH hypothesis. Besides, the effect of political uncertainty is sector-specific. While some sectors emerged winners (healthcare, oil and gas, real estate, defense, financials and consumer goods and services), others took the opposite route (technology and utilities). The winning industries are generally those that will benefit from the new administration's focus on rebuilding infrastructure, renegotiating trade agreements, reforming tax policy and labour laws, increasing defense funding, easing restrictions on energy production, and rolling back Obamacare.

[†] CATT, University of Pau; *Corresponding Author:* Avenue du Doyen Poplawski, 64000 Pau, France; Tel: +33 (0) 5 59 40 80 01, Fax: +33 (0) 5 59 40 80 10, E-mail: jamal.bouoiyour@univ-pau.fr

[‡] University of Tunis; University of Pau; E-mail: s.refk@yahoo.fr

# Graphical abstract

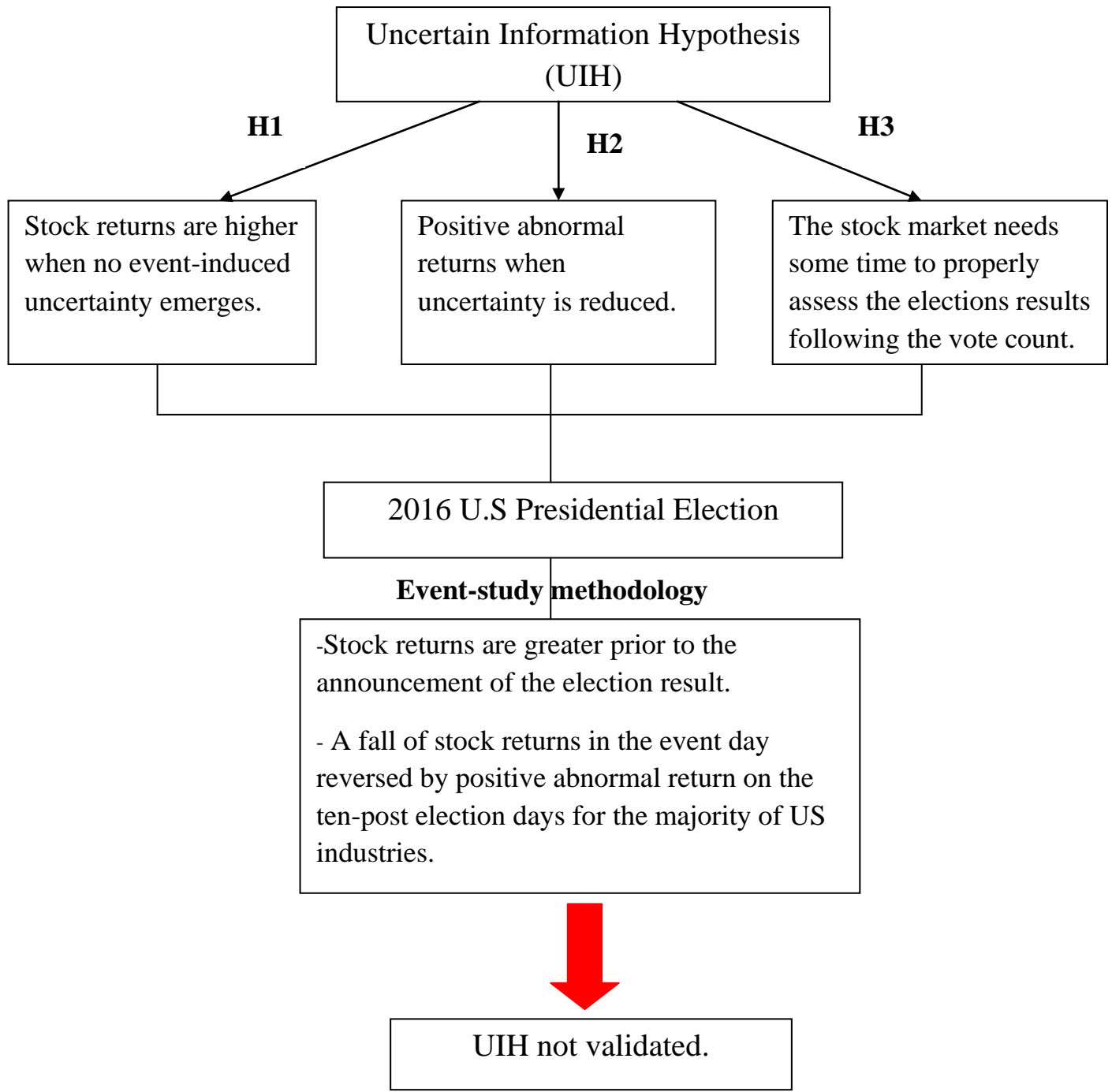

# 1. Introduction

The political uncertainty is a prevalent phenomenon which is immanent to the political process. Within the political science literature, the political uncertainty refers to the lack of assuredness. Dahl et al. (1963) indicated that elections, wars, governmental processes and threats are all viewed as uncertain political phenomena. Even though political uncertainty takes various shapes and forms including changes in the government and changes in the domestic and foreign policies, the present research focuses on one kind of political uncertainty, which is associated with elections. The latter constitute a major event for re-distribution of political power, which may have meaningful implications for the future political and economic prospects of a country. The political uncertainty naturally emerges since different candidates running for office, if elected, will undertake different policies, and election outcomes are uncertain. One can assert that political uncertainty is just a reflection of policy uncertainty. These two forms of uncertainty, while heavily associated, have different characteristics. Policy uncertainty is the uncertainty with respect the government policies (macroeconomic, monetary and fiscal policies) and their effects on the economic development and financial markets (Pasquariello 2014). The political uncertainty, nevertheless, encompasses both uncertainty about the election result and uncertainty regarding the policies that may ensue from that outcome (Pasquariello and Zafeiridou 2014).

The existing literature documents that the political uncertainty particularly, may exert a significant effect on both the returns and the risk levels of financial assets (Pantzalis et al., 1999; Nippani and Medlin, 2002; Li and Born, 2006; He et al., 2009; Jones and Banning, 2009; Sy and Al Zaman, 2011; Goodell and Bodey, 2012). For example, Pantzalis et al. (1999) assessed the responses of stock market indices across 33 countries to political election dates during the sample period from 1974 to 1995. They claimed that political uncertainty falls over the two weeks prior to elections, yielding to a rise of stock market valuations, consistently with the uncertain information hypothesis of Brown et al. (1988). Nippani and Medlin (2002), Nippani and Arize (2005), Li and Born (2006), He et al. (2009), and Goodell and Bodey (2012) examined the impact of US presidential elections on stock markets, and deduced that the ongoing uncertainty over the elections is reflected in the behavior of stock prices. Using polling data on the US presidential elections for the sample period from 1964 to 2000, Li and Born (2006) showed that stock prices climb prior to the presidential elections when the election result is uncertain. Nippani and Arize (2005), and He et al. (2009) investigated the reaction of stock markets to the delayed result of the 2000 presidential election, and found that stock markets are negatively influenced by the uncertainty surrounding the election outcome. Goodell and Bodey (2012) indicated that a collapsing uncertainty around US presidential elections prompts a drop in stock prices. Goodell and Vahamaa (2013) studied the impacts of political uncertainty and the political process on implied stock market volatility during US presidential election cycles. They found that the relationship between implied volatility and the election probability of the eventual winner is positive.

Generally speaking, the financial markets tend to react to new information with respect political events that may exert a significant influence on the country's macroeconomic, fiscal and monetary policies. In fact, the political events are followed by investors who form or revise their expectations based on the results of these events. Informational efficiency hypothesis assumes that markets absorb news and political trends into asset prices in anticipation of election results. Much of the uncertainty surrounding the outcome may be resolved prior to the election date. Such policy changes are typically associated to a decrease of stock prices, particularly if the uncertainty is greater (Pastor and Veronesi 2012; Bouoiyour and Selmi 2016). Once the political uncertainty is mitigated, stock prices would rise again (Pantzalis et al. 2000). Brown et al. (1988) argued that as uncertainty is reduced, price changes tend to be positive on average. On the contrary, if the election outcome does not permit market participants to immediately and effectively evaluate the effect on the nation's future, then the election result constitutes an uncertainty inducing surprise. In this case, positive price changes should be anticipated after the election, i.e. until uncertainty about the policies to be achieved by the winner is resolved.

Against all odds, polls, and projections, the Republican candidate -Donald Trump- claimed win in the 2016 US presidential election, defeating Democratic nominee Hillary Clinton. Financial markets had widely priced in a victory for Clinton, who they viewed as a better short-run outcome because she represented few unknowns and thus less uncertainty. This study examines, at sectoral level, the US stock market behavior around the 2016 presidential election and addresses the following questions. Do markets anticipate the election outcome? Are US stock markets efficient? To what extent does election outcome resolve uncertainty? Are stock markets resilient in dealing with the uncertainty arising from political shocks? Is there homogeneity in stock market behavior around the US election results between the different sectors? We explore these questions using a standard event study methodology that examines abnormal return behavior around the election date. The study tests also the uncertain information hypothesis (UIH) of Brown et al. (1988).

By gathering separately the responses of eight of largest firms in the Dow Jones, S&P500 and Nasdaq Composite indices, we document that while the Trump's win in US presidential elections has played a negative role on the abnormal return in the day event, a positive reaction for almost all sectors was found during the post-election period (except technology and utilities). These findings are not consistent with uncertain information hypothesis of Brown et al. (1988). Moreover, the effect of political uncertainty around elections is likely to be sector-specific. In particular, the uncertainty surrounding the Trump's win divides the US stock markets into winners (health care, oil and gas, real estate, defense, financials and consumer goods and services) and losers (utilities and technology). Several elements of explanations have been offered to explain the heterogeneous reactions of U.S. companies.

The remainder of the paper is organized as follows. In Section 2, we present our formal hypothesis and describe the methodology and the data sources. Section 3 describes the empirical findings, while section 4 checks their robustness. Finally, in Section 5, the conclusions of the analysis are summarized and policy implications are discussed

## 2. Testable hypotheses, methodology

Since Efficient Market Hypothesis (EMH) has arisen in the 1960s (Fama, 1965, 1970; Samuelson, 1965), it has been subject to a huge number of researches. Under the assumption of rational investor, this hypothesis postulates that share prices completely reflect information and expectation, and that any new information is incorporated into equity prices very quickly. In contrast, empirical studies showed that stock prices do not often fully reflect all information. This contradiction has yielded to the appearance of new hypotheses in behavioral finance including the Uncertain Information Hypothesis (UIH) of Brown et al (1988). The Uncertain Information Hypothesis assumes that anxiety will rise in financial markets following the occurrence of unexpected event. So that investors cannot appropriately respond to unanticipated new information and thus they could in the early stages set security prices below their fundamental values. Moreover, the UIH asserts that the stock return is likely to be greater than the average return over periods when no event-induced uncertainty happens. When election-induced uncertainty is mitigated, we expect a positive response of the cumulative abnormal returns (CARs) in the time period following the election. The first hypothesis to be tested throughout this study consists, therefore, of two parts:

$$H1_a : CAR_{-10;0} \succ 0 \tag{1}$$

$$H1_b : CAR_{0;+1} \succ 0 \tag{2}$$

One can expect also that the election results partly resolve the great uncertainty surrounding the unanticipated political event and that the market requires more time to adequately evaluate the elections' consequences following the vote count. If there is a greater uncertainty resolution following the election outcome, we would notice post-election positive abnormal returns. In brief, the UIH assumes that a mitigation of uncertainty prompted positive observed returns and that wider uncertainty reduction typically leads to greater observed returns. In this study, we investigate the ten-day period after the election date to test our second hypothesis:

$$H2: CAR_{+1;+10} \succ 0 \qquad (3)$$

The purpose of this study is to examine stock market behavior around political election dates. We focus our analysis on the 2016 US presidential election outcome and investigate the impact of the Trump's win on different US industries. The final result of the election was disclosed on Tuesday 08 November 2016, which we subsequently view as the announcement day. Our sample data include eight sectors of three US stock price indices: The Dow Jones Industrial Average tracks the prices of 30 widely-traded stocks on the New York Stock Exchange. This is the most known stock market index in the world but it is not representative of the market as a whole. The Nasdaq Composite is the market capitalization-weighted index of approximately 3,000 common equities listed on the Nasdaq stock exchange. The Standard and Poor's 500 (S&P 500) Composite Stock Price Index covers the performance of 500 largest capitalization stocks. For each index, the selected companies include financials (banks, insurance, reinsurance and financial services), oil and gas (oil and gas producers, oil equipment and services), real estate, consumer goods (household goods, home construction, personal goods and tobacco) and services (retail, media, travel and leisure), defense, pharmaceuticals, technology (software and computer services, and technology hardware and equipment) and Utilities (electricity, gas, power generation and water). Each sector index represents a capitalization-weighted portfolio of the largest S&P 500 companies in this sector. The data of sectoral Dow Jones Industrial Average, S&P 500, Nasdaq stock indices are available at Datastream database.

We employ the standard market model event study methodology as depicted by Dodd and Warner (1983) and Brown and Warner (1985). Before presenting the conducted procedure, we should point out that an event studies investigates the average stock market response to a specific stock market event, by averaging among the same event in different companies. The best findings with an event study are revealed when the exact date of the event is known or identified. We define the day "0" as the day of the event for a given equity. Thereafter, the estimation and event windows can be determined (Figure 1). The interval [T1+1, T2] is the event window with length L2=T2-T1-1, whereas the interval [T0+1, T1] is the estimation window with length L1=T1-T0-1. The length of the event window often depends on the ability to accurately date the announcement date. If one is able to date it precisely, the event window will be less lengthy and capturing the abnormal returns will be more proper and effective. We should mention here that the length of the event window including the event announcement days normally ranges between 21 and 121 days (Peterson 1989).

**[Insert Figure 1 here]**

For our case of study, we use for each sector a maximum of 120 daily stock return observations for the period around the ultimate election result, beginning at day - 115 and ending at day + 5 relative to the event. The first 105 days (- 115 through -10) is denoted as "the estimation period", and the following 21 days (- 10 through + 10) is designated as "the event period". The cumulative abnormal return (CAR) for a sector $i$ during the event window $[\tau_1 ; \tau_2]$ surrounding the event day $t = 0$, where $[\tau_1 ; \tau_2] = \in [-10 ; +10]$, is expressed as follows:

$$CAR_{i,[\tau_1,\tau_2]} = \sum_{t=\tau_1}^{\tau_2}(R_{i,t} - \hat{\alpha}_i - \hat{\beta}_i R_{M,t}) \tag{4}$$

where $CAR_{i,[\tau_1,\tau_2]}$ is the cumulative abnormal return of share $i$ during the event window $[\tau_1; \tau_2]$, $R_{i,t}$ is the realized return of stock $i$ on day $t$[3], $R_{M,t}$ is the return of the benchmark index of sector $i$, $\hat{\alpha}_i$ and $\hat{\beta}_i$ are the regression estimates from an ordinary least squares (OLS) regression for 105 trading day estimation period until $t = -10$. We utilize the Datastream's value-weighted total return stock market index of sector $i$'s country of origin as the benchmark index. We set our event day for the Trump's victory event to Tuesday 08 November 2016.

We apply, then, a regression analysis to identify the determinants of the observed cumulative abnormal return for each sector. The OLS regression to be estimated is denoted as:

$$CAR_{i,[\tau_1,\tau_2]} = \delta_0 + \delta_1 Trump + \delta_2 Size + \delta_3 Income + \varepsilon_i \tag{5}$$

---
[3] Daily stock returns are calculated as the first natural logarithmic difference of the underlying stock price.

where $CAR_{i,[\tau_1,\tau_2]}$ is the dependent variable, *Trump* is a dummy variable which takes the value of one on the first day of trading after the US election outcome and zero otherwise, *size* is the logarithm of the total assets of a company in U.S. dollars in the year prior to the event, and the *Income* is the logarithm of the net income of a company in dollars in the year prior to the event, and $\varepsilon_i$ is the error term. The explanatory variables "*size*" and "*Income*" were chosen based on recent event studies showing that the largest companies are more threatened by sudden events or political changes, and the response of stocks to uncertainty surrounding an event may depend on the net income of a firm in the year before the occurrence of the event (Kolaric and Schiereck 2016; Bouoiyour and Selmi 2016).

3. **Empirical results**

In this section we present the event study results. We begin the analysis related to the three hypotheses of Brown et al. (1988) by depicting the performances of the cumulative abnormal return for different sectors of three US stock price indices (Dow Jones Industrial Average, S&P 500 and Nasdaq) around the day relative to announcement of Trump's victory on 08 November 2016 (t=0) and for different event windows: [-10; 0], [0; +1] and [0+10]. The first hypothesis drawn from the UIH of Brown et al. (1988) states that the tock returns are higher when no event-induced uncertainty emerges. The second hypothesis assumes that positive abnormal returns when uncertainty is reduced (i.e., when the election result is announced or becomes certain), while the third hypothesis postulates that the stock market requires time to properly evaluate the elections results following the vote count. Precisely, greater positive returns should be associated with greater reductions in uncertainty

Figure 2 indicates that US stock markets' responses to the election outcome is not uniform across industries either for the announcement day or the [−10; + 10] event window. In other words, while all companies face increasing uncertainty, the Trump's win had varying sectoral effects. Ten days prior to the election vote (i.e., the event period -10; 0), positive abnormal returns are found for all U.S. companies under study. A sharp decrease in stock values surrounding the election result (i.e., the event day 0; 0) is later reversed by a jump in share prices on the next day (i.e., the event period 0; +1). Potentially, the win of Donald Trump is associated to severe stock prices declines for all the sectors on the day relative to the announcement of US election results (t=0). However, we show that the majority of sectors rebounded. The effect of political uncertainty is positive in the ten days after to the vote count (i.e., the event period 0; +10). This holds true for the three

US stock indices under study. Exceptionally, for utilities and technology, we observe positive abnormal returns in the ten days before the release of the event (i.e., the event period -10; 0), negative reactions since the day relative to the announcement of the election result (t=0) and the ten-days post election. The findings of the event study of the CAR performances around the 2016 U.S. presidential election are not in line with the UIH hypothesis. Although previous studies indicated that anxiety around political events might have detrimental effect on stock returns, the impact of Trump's win in U.S. presidential election is surprising. In general, policy changes are followed by collapsing stock prices, especially when the uncertainty is strong. Once the political uncertainty is reduced, positive changes in stock returns are highly expected (Pantzalis et al. 2000). By delving into the case of 2016 US presidential election, a sharp decrease of abnormal returns were seen for all the U.S. companies in the day relative to the announcement of the election outcome (t=0), before surging again after the vote count. While the new administration' policy directions remain unclear, the investors have bet the newly US president will deliver on some of his most basic campaign promises including the improvement of infrastructure spending and cutting corporate taxes.

**[Insert Figure 2 about here]**

Table 1 summarizes the event study of the cumulative abnormal returns of different U.S. companies around the 2016 U.S. presidential election day and the post-election period. We attempt from a comparative analysis between the event day [0; 0] and the event window [+1; +10] to test the UIH hypothesis postulating that the election outcome partially resolve prior uncertainty and the stock market requires time to assess the elections' effects following the vote. The findings reported in Table do not appear consistent with the UIH hypothesis; for the majority of U.S. industries studied, there is a negative market reaction in the event day, and the effect becomes positive through the ten-day period following the election outcome. In particular, we show that the announcement of Trump victory (i.e., the event day [0; 0]) resulted in statistically significant negative CARs, being somewhat stronger for utilities, technology, oil and gas, financials and defense (in this order) than for consumer goods and services, real estate and health care. Overall, it appears that Donald Trump's win had market-wide repercussions, leading to a decline of all the companies for the three considered US stock markets, but the collapse of healthcare share prices (in particular) is not as severe. The same sectors which struggled after the election show positive reactions during the [+ 1;

+10] event window, except utilities and technology which appear more damaged during the post-election period (i.e., negative responses).

Financials reacted rapidly and positively to Trump's win; thus, this sector ended higher after starting the session (day event [0; 0]) with sharp losses. One of the major causes for the jump seems to be because Donald Trump is expected to lessen regulation hampering bank profitability.

Also, the response of oil and gas market bounced back after the presidential election outcome ([+1; +10] event window) as Trump declared his desire to revive the energy sector. The ultimate US election result comes as good news for both crude oil and natural gas due to President-elect Trump plans to minimize regulatory restrictions on crude and gas exploration. In addition, the new Trump's administration will benefit the fossil fuel business and independent oil and gas drillers, promising few regulations on issues such as methane emissions from oil and gas drilling, ozone rules and renewable fuels, and higher access to federal lands. Furthermore, Trump has expressed displeasure for alternative forms of energy, describing them as expensive and needing largest subsidies to work appropriately. In this context, the Trump administration stated that it would reform all forms of energy while trying to reflect their true costs. However, despite these fruitful promises, the reaction of this sector to Trump's win seems weaker. This may be attributed to Donald Trump's aggressive stance towards Mexico -a main partner in the American energy industry- that could severely harm US oil and gas exports south of the border. We should mention at this stage that US gas imports to Mexico exceeded Mexican domestic production in 2016. Also, the renewable energy industries palpitated at the prospect of less commitment to reforms that unhurried climate change.

Differently, real estate does not react negatively to the announcement of Trump victory as the rest of sectors. The US election outcome exerts a positive influence on the housing sector during the day event and the [+1; +10] window event, even if we note a slight increase after the election results. Not surprisingly, for the first time in history, home builders and real estate businessman see one of their own becoming the elect-US president. They are optimistic about Donald Trump stimulating this sector, in the form of lower tax rates or enhancement of roads, bridges, public transit and wider infrastructure spending. Nonetheless, Trump's eloquence on immigration could concern housing investors in big cities such as New York and San Francisco. With Trump's "America First" approach to alienating partners abroad, America will become more isolated and less open, which could seriously impede the international demand for US luxury housing since the foreign buyers constitute a large part of the real estate market.

The consumer goods and services sector is also one of the winners from Trump victory. While its response to the announcement of Trump's win was negative, it bounced back after the event day. This reflects a rise in the consumer confidence[4], showing that Americans became more optimistic about their finances and the economy after Trump victory. Nevertheless, some of the Trump's proclamations during his campaign exacerbated doubts about globalization and some trade deals, resulting more expensive imported products due to excise taxes that could unhurt consumer goods.

Further, our findings indicate that defense sector is one of the winners from the Trump's presidential win. While the announcement of the election outcome had first affected negatively (but moderately) this sector, we notice a positive response of defense firms during the post-election period as investors in this sector believe that they would post larger benefits under Trump presidency. We can attribute this result to the new administration promises to increase the size of the Army and the Marine Corps, build newly ships for the Navy and to overhaul the aerial warfare service branch, and modernize the nuclear arsenal.

Our results reveal that the health care is the biggest winner from Trump victory given to its heavier support of the pharmaceutical sector and because the drug pricing reforms, proposed by Hillary Clinton's campaign, seem unlikely to materialize. In brief, during [+1; +10], pharmaceutical shares are likely to bounce upward as investors expected relief from the stronger scrutiny of drug prices. Indeed, the health-care industry would gain from the Affordable Care Act; more people bought insurance and had better access to medical care.

However, utilities and technology seem the most damaged from Trump victory. Utilities, especially those levered to natural gas, solar and other renewables, dropped markedly following the victory for Donald Trump. This may mainly due to the Trump's condemnatory proponent of punishing those firms who move facilities out of America, in particular to Mexico. Regarding the technology sector, the Trump campaign made little outreach to issues influencing the tech industry. This little interest may be contradictory with the Trump's campaign message to spur US economic growth. In fact, the tech industry accounts for 12 percent of all jobs, according to the US Bureau of Labor Statistics, and thus the neglect of effective technology policies will have detrimental impact on America's economic development and competitiveness. Further, the Trump's opposition to H1B visas[5] for high-skilled immigrants will harm substantially the capability of US tech firms to hire the engineers, data scientists and the information technology workers they need from other countries.

---

[4] The University of Michigan claimed that the index of consumer sentiment increased from 87.2 in October to 93.8 in the post-election period.

Moreover, the size of the firm is likely to exert significant and negative influence on all the U.S. stock market sectors and across the [0; 0] and [+1; +10] event windows, highlighting that biggest companies are likely to be more harmed by the uncertainty around the election results. The profits of U.S. firms do not help to consistently explain the Dow Jones, S&P 500 and Nasdaq evolutions, as the net income exerts a weak and positive influence on limited sectors (financials and oil and gas for Dow Jones, oil and gas and real estate for S&P 500, and financials, oil and gas and technology for Nasdaq).

**[Insert Table 1 here]**

4. Robustness

There exist different ways to ascertain whether our results are fairly solid. In this study, to check the robustness of our findings, we have tested their sensitivity to the inclusion of further control variables. In general, global financial and economic factors could be channels through which fluctuations in the world's economic and financial conditions are transmitted to the different sectors of US stock markets. These factors include the US volatility index (VIX), and the world gold price (gold). Supplementary control variables have been incorporated including silver and Bitcoin prices. The precious metals (gold and silver) have been largely perceived as a hedge against sudden shocks and also a safe haven over extreme stock market fluctuations. In the present study, we tried to see if US investors still rush to precious metals over the announcement of US presidential election results or if they get scared to seek out gold and silver. According to Baur and McDermott (2010), we characterize safe havens by their negative and significant correlations with asset markets during financial turmoil or troubled times. Moreover, the literature in finance field has been frequently relied on proxies of uncertainty, most of which have the advantage of being directly observable. Such proxies include the implied volatility of stock returns (i.e., VIX). The interest here is to use an index that reflects more adequately the great anxiety over U.S. presidential election. The volatility index is a sentiment indicator that allows determining when there is too much optimism or pessimism in the market. Also, we should point out that VIX responds sensitively to all events (reflecting both economic and geopolitical issues) that may cause uncertainty, and the Trump's win is no exception. Overall, it helps reaching further insights about how the stock markets react to global market news. The Bitcoin is a relatively new phenomenon created in 2009. It is a peer-to-peer network that allows the transfer of ownership without the need of a third party. Bitcoin is regarded as the best-known digital currency to date. Although some consider Bitcoin to be a major financial innovation in recent years (Ciaian et al. 2014; Bouoiyour et al. 2016), others suggest that the excessive volatility observed in this market is a major concern (Yermack 2014). The Bitcoin's climb alongside the announcement of Trump's victory has led some to proclaim it as a "digital gold" and affirm its validity as a safe haven investment.

---

[5] H1B visas are designed to allow US employers to recruit foreign professionals in specialty occupations within the America for well specified period of time.

In brief, the equation to be estimated is denoted as:

$$CAR_{i,[\tau_1,\tau_2]} = \lambda_0 + \lambda_1 Event + \lambda_3 Size + \lambda_4 Income_t + \lambda_5 VIX_t + \lambda_6 Gold_t + \lambda_7 Silver_t + \lambda_8 Bitcoin_t + \xi_i \quad (6)$$

where $CAR_{i,[\tau_1,\tau_2]}$ is the cumulative abnormal returns and $\xi_i$ is the error term.

The results are reported in Table 2. We show that the consideration of additional control variable have not fundamentally changed our findings for the three stock price indices studied; We robustly do not support the uncertain information hypothesis of Brown et al. (1988). We usually note that a drop of abnormal stock returns around the election day event (t=0) is reversed latter by positive abnormal return responses (i.e., event period +1; +10), except for utilities and technology where we show that the cumulative abnormal return reactions are negative. Also, we unambiguously document that the announcement of the Trump's win in 2016 US election exerted a varying effects across US companies. Specifically, it divided the US markets (in particular, Dow Jones, S&P 500 and Nasdaq) into losers (technology and utilities) and winners (health care, oil and gas, real estate, defense, financials and consumer goods and services). The size of company affects negatively the US stock market sectors, sustaining the evidence that largest firms are more exposed to uncertainty surrounding Trump's presidency than smallest companies. The profits of US firms do not exert strong impact the performance of the companies. The implied volatility index has a negative influence on the different sectors of US stocks, indicating that the stock returns decrease as the VIX increases. In addition, the precious metals (gold and with less extent silver) have a negative influence on the abnormal cumulative returns for almost all the industries. Typically, when the economy witnessed an evolving volatility that may impede shares' valuation, investors may shift their funds from stocks and invest them in the gold and silver markets until the economy rebounds. In this context, precious metals could act as a stabilizer control in investment portfolios, and play as safe haven during turbulent times (Baur and Lucey 2010). Besides, Bitcoin price is likely to have a negative and significant impact on US companies. Remarkably, the effect of Bitcoin on stocks seems more pronounced than that of gold and silver. Although Bitcoin spikes after the announcement of the US election outcome spotlights a new confidence in Bitcoin as a safe haven, investment professionals have been heavily reluctant to give this nascent crypto-currency such status. Given the great anxiety over Trump's victory, it is obvious that investors will try to seek an easy and secure alternative. Our results suggest the ability of Bitcoin, gold and silver (in this order) to act as a safe haven during uncertain periods. Nevertheless, dubbing Bitcoin a safe haven obfuscates the fact that bitcoin is a high-risk, volatile and speculative investment.

**[Insert Table 2 here]**

After the announcement of the presidential election outcome, it was anticipated whose companies were poised to gain. Healthcare, housing builders, oil and gas, defense and financial industries would generally behave well. Trump proposed a tax-free health savings account to allow individuals to save money to pay for healthcare costs in order to subtract the cost of their premiums for tax purposes. Trump also urged the prominence of price transparency from all providers (for instance, clinics, hospitals and other healthcare organizations) and raising competition while attempting to minimize insurance costs and stimulate consumer satisfaction. For oil and gas firms, It is clearer that fossil fuels in the United States pledged to make the United States "energy independent" by allowing access to new areas of the country, including federal land, to oil and gas development, and revising environmental and climate policy standards, and completely removing subsidies to renewables. Moreover, protecting the U.S. borders was one of the main Trump campaign' focus. Therefore, surging defense funding will be top priorities for the administration, which will undoubtedly exert a positive influence on the defense industry. For financial sector, there is an optimism that there will be a lighter regulatory hand, but the behavior of financial and banking companies would still conditional on the state of the economy. At the same time, the biggest firms involved with technology and utilities would see stocks slide. Apple and Amazon have been largely criticized by Donald Trump; the first for making iPhones in China, and the second for disobeying antitrust laws.

Also, Silicon Valley has long vigorously defended an expanded H-1B Visa programme to allow access to highly skilled workers. Some technology executives think a Trump administration may impede innovation by opposing these visas. These circumstances will hinder the capability of U.S. tech firms to hire the engineers, data scientists, as well as the information technology workers they need from other countries, and then moisten start-ups and damage projects in both the private sector and in the federal government. With wider utility investments in plants, pipelines and other infrastructure, the current investments will build the power generation mix for the next years. Under the Clean Power Plan aimed at fighting against the global warning and the greenhouse gases emissions -the most challenging problems of the world- in accordance with standards set by the Paris climate accord, U.S. utilities are opting to substitute retiring coal plants with wind and solar facilities. However, President-elect Donald Trump has vowed that when he is inaugurated he will kill the Clean Power Plan and pull out of the Paris Agreement, two pillars of the Obama administration's drastic efforts to battle against climate change by mitigating greenhouse gas emissions.

Beyond the political fights, the Trump's promises to Trump has proposed to lessen the corporate tax rate to 15 per cent (from 35 per cent) to stimulate start-ups, develop existing companies, promote fund in corporate infrastructure, and make the country more competitive tax environment for multinational corporations. Immigration was another disputable Trump's campaign issue. Based on the Bureau of Labor Statistics, a fully enforcing immigration law would recoil the labor force by 11.2 million workers. In addition, the president-elect put forward to change the North American Free Trade Agreement (NAFTA) would be economically harmful, interrupting investment continuity for industries. Certainly, if the United States were to unilaterally impose temporary trade restrictions, other countries may react in kind with punitive restrictions of their own, which would put a hindrance on global trade and lead the way on menacing protectionism. In this context, many international organizations (in particular, the International Monetary Fund and the World Trade Organization) are worried that the withdrawal from NAFTA, the renegotiation free-trade agreements resulting more isolated and less open US markets would cause a trade slowdown that would damage the global economy. A repeal of trade agreements would prompt shortages in raw materials climbing the prices and adversely influencing the availability of consumer products. But this remains conditional to the overall congress opinion and the legal challenges from private firms which may play a pivotal role in deterring Trump's administration from implementing these measures. As policy directions clarify over time, U.S. industries can firm up their reactions depending to the resulting changes in the operating environment.

## 5. Conclusions

This paper examines the effects political uncertainty on stock market performance around the 2016 US presidential election. Previous studies have documented that political election are heavily associated with periods of considerable public uncertainty, and therefore, it is of interest to empirically assess the effects of election-induced uncertainty on stock market. The study examines a sample of U.S industries for testing the uncertain information hypothesis. It focuses on market reaction to announcements of new political event using the event analysis methodology.

Our results reveal that Trump's win had a significant impact on the valuation of companies for variety of US stock price indices (Dow Jones Industrial Average, S&P 500 and Nasdaq Composite). However, the findings of the study do not provide a consistent conclusion regarding the existence of uncertain information content hypothesis in the U.S. stock market. While prior research on the effects of changes in government policy showed a negative influence on equity markets, the effects of Donald Trump swept to victory on US stocks is unanticipated. Normally, companies have to make prominent choices based on the expected future economic policy decisions of the new government and the resulting policy circumstances (Brogaard and Detzel 2015; Schiereck et al. 2016). In this way, the Trump's win can be viewed as a drastic change in government policy. Such policy changes are typically linked to a drop of stock prices, particularly if the uncertainty is greater (Pastor and Veronesi 2012). Once the political uncertainty is mitigated, stock prices would rise again (Pantzalis et al. 2000). In the case of 2016 US presidential election, investors and traders who some days prior to the election saw a Donald Trump victory as the heaviest downside risk to the stock market, are now embracing the outcomes. After an initial notable collapse during the event day (t=0), stocks rallied after the vote count, with investors making quick recalculations on various sectors. While many Trump policy proposals are still vague and ill-defined, investors are betting that the Trump's promises will recharge the US economy by cutting taxes, rolling back regulation and boosting infrastructure spending. In other words, the basis for the rally is hopefulness about altering Obamanomics consisting of increasing taxes and improving regulation. Also, the Trump's zero-sum approach or "America is first" -in favour of isolationism- to encouraging investments at home and antagonizing partners abroad exerted a positive effect on stocks.

While all of the U.S. companies face a great political uncertainty around U.S. presidential election, varying responses were found. In particular, the Trump victory divides the U.S. stock markets under study into two main groups: (1) a group of winners which is formed by financials, oil and gas, real estate, consumer goods and services, defense and health care, and (2) a group of losers which contains utilities and technology. Part of this division can be explained by the Trump campaign promises to ensure an economic environment of lowered regulation, reduced global trade, increased infrastructure spending and a cancellation of Obamacare and climate policies.

# Figure 1. Event study windows

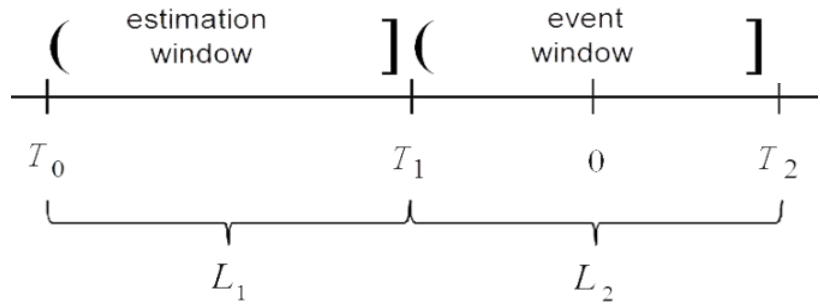

# Figure 2. Cumulative abnormal return of US stock indices by sector: [−10; +10] event window

| Dow Jones Industrial Average | S&P 500 | Nasdaq Composite |
|---|---|---|
| Financials | | |
| 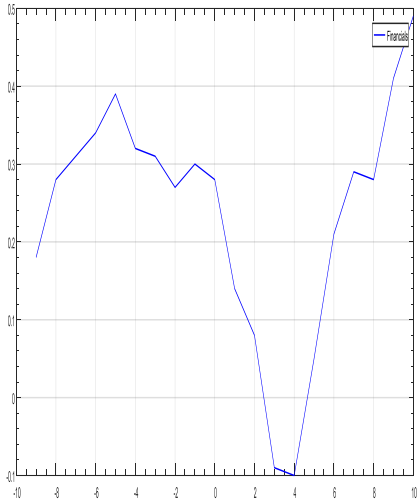 | 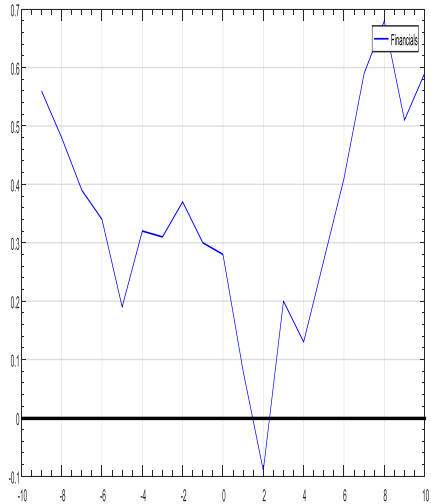 | 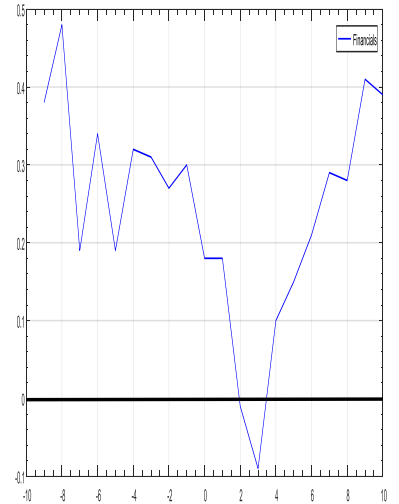 |
| Oil and gas | | |
| 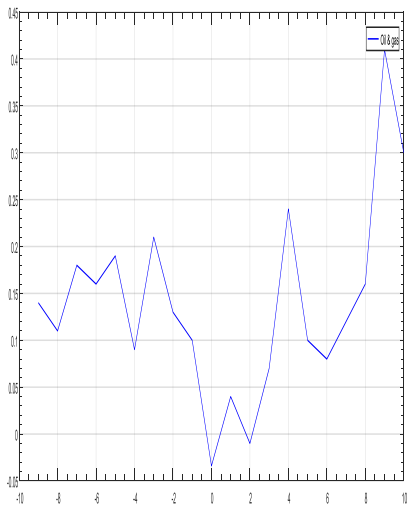 | 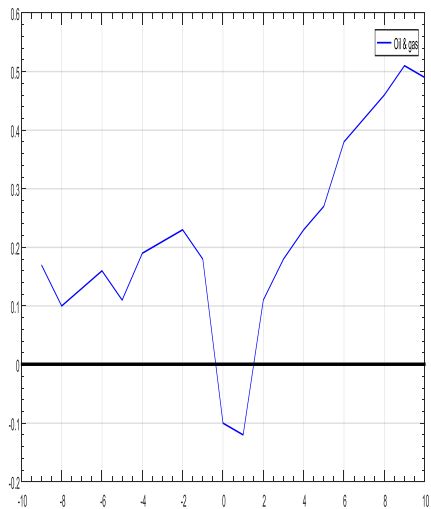 | 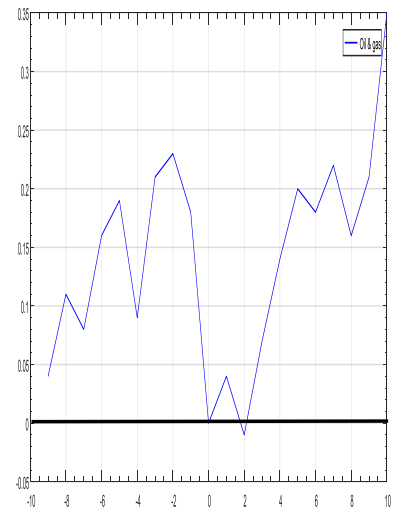 |

## Real estate

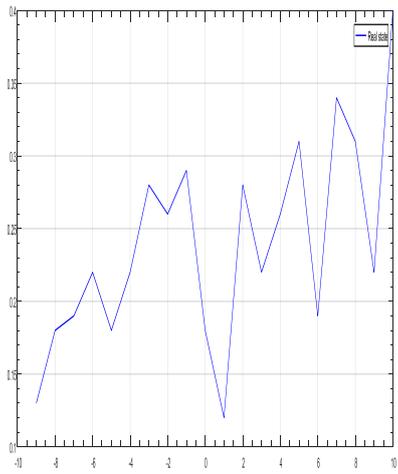 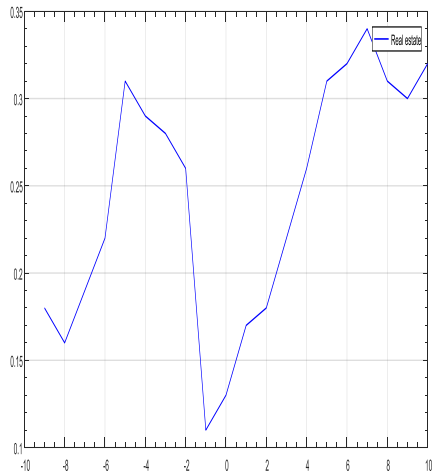 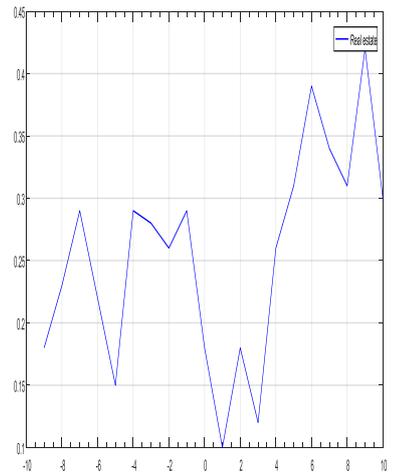

## Defense sector

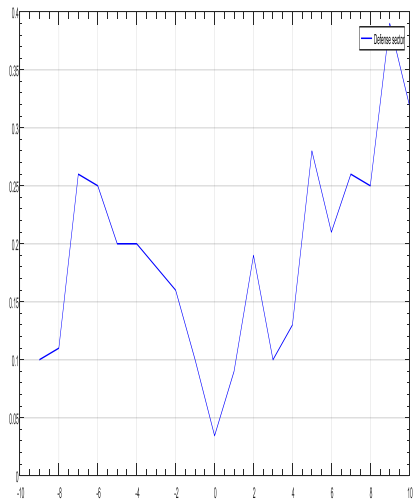 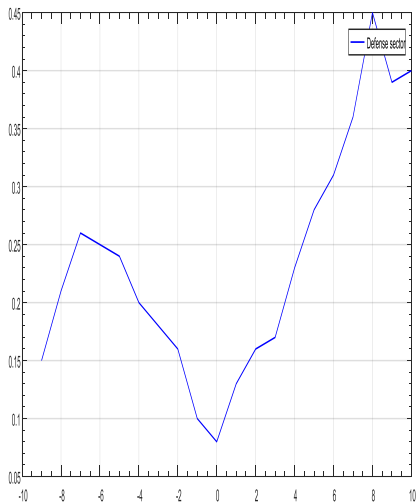 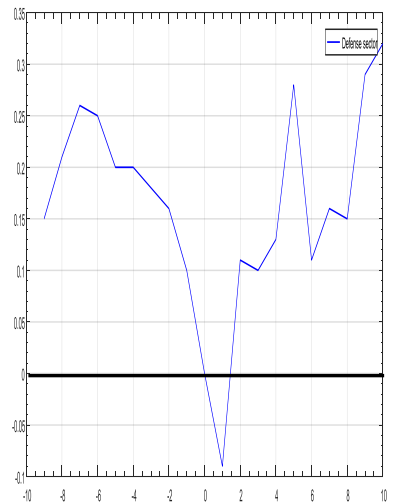

## Health care

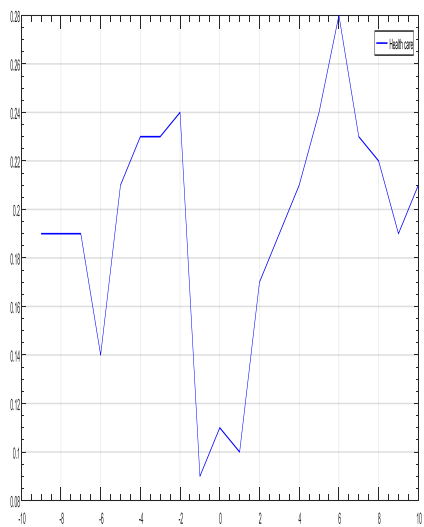 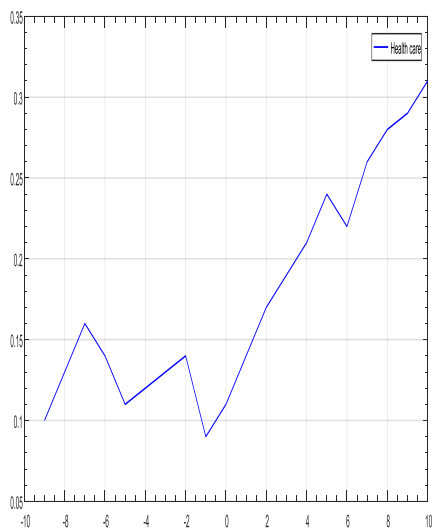 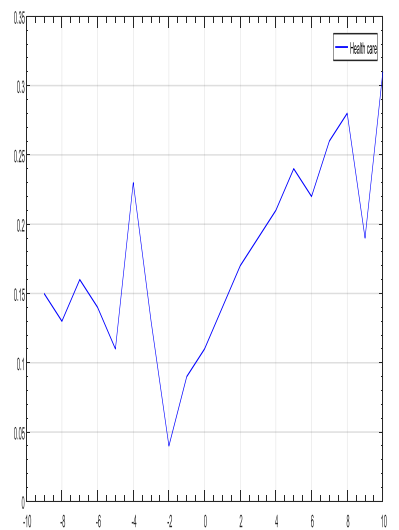

## Consumer goods and services

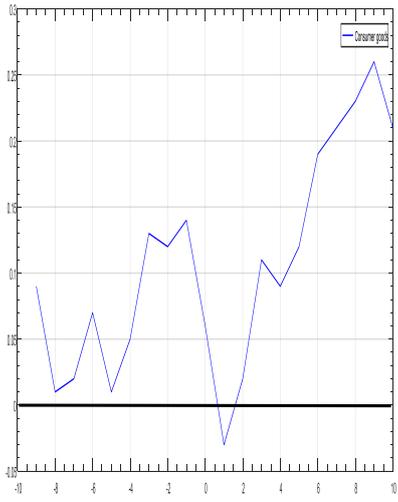 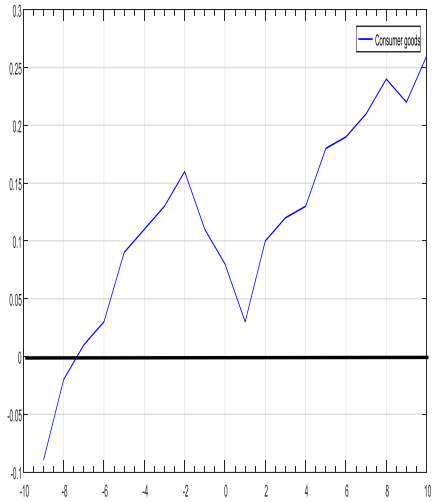 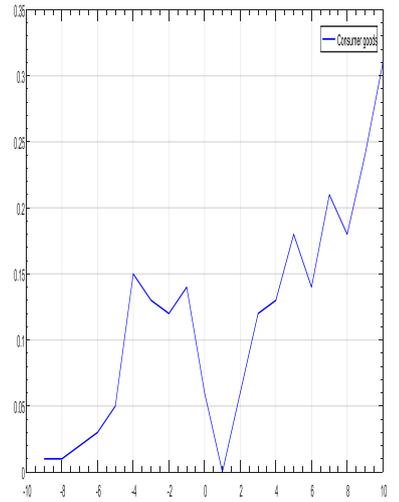

## Technology

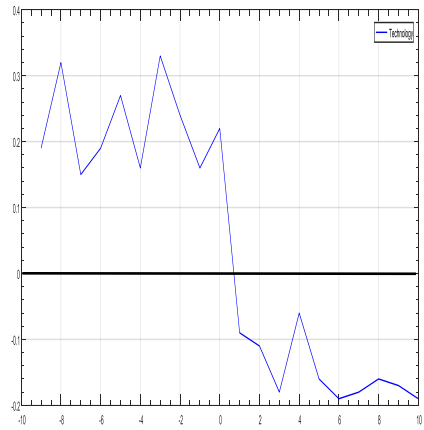 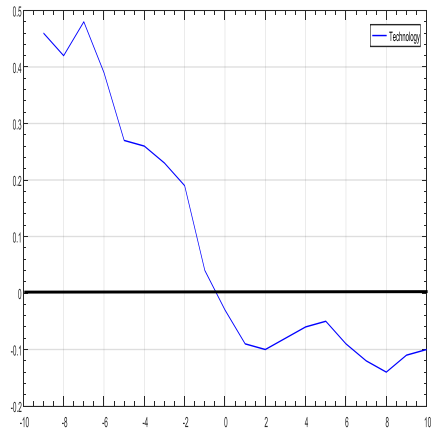 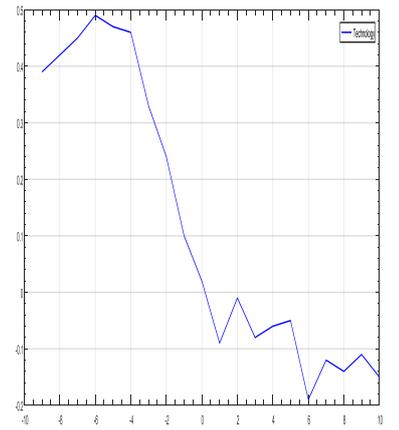

## Utilities

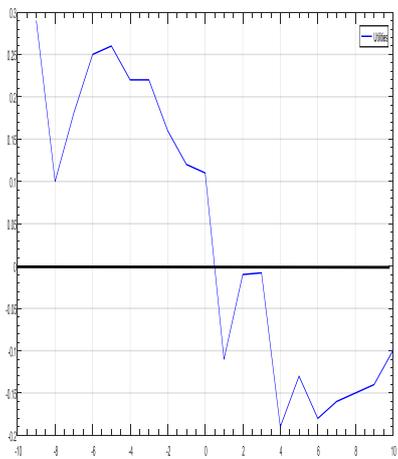 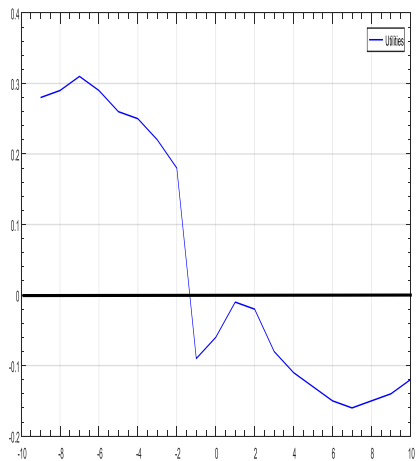 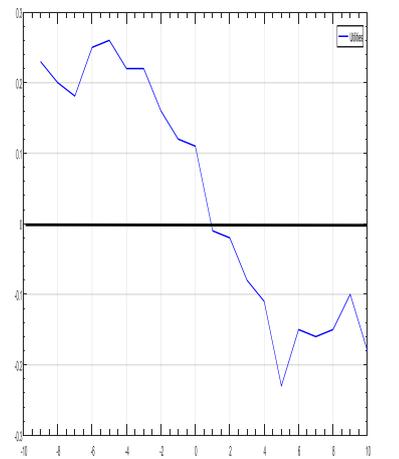

**Table 1. Sectoral impacts of 2016 U.S. presidential election on U.S. stock markets**

|  | Financials | Oil and gas | Real estate | Consumer goods & services | Defense | Health care | Technology | Utilities |
|---|---|---|---|---|---|---|---|---|
| **Dow Jones Industrial Average** | | | | | | | | |
| *Event day [0 ; 0]* | | | | | | | | |
| Constant | 0.347377 (0.2900) | 0.450940* (0.0465) | 0.227745* (0.0806) | 0.565629 (0.9331) | -0.01710 (0.9819) | 0.391338* (0.0315) | 0.451239* (0.0616) | 0.290433 (0.5893) |
| Trump | -0.18336* (0.0991) | -0.196*** (0.0008) | 0.08427* (0.0929) | -0.104** (0.0010) | -0.0220* (0.0314) | 0.09586* (0.0527) | -0.24193** (0.0042) | -0.2797* (0.0298) |
| Size | -0.13556* (0.0799) | -0.10943* (0.0634) | -0.08303* (0.0309) | -0.028** (0.0073) | -0.1424* (0.0497) | 0.454829 (0.2674) | -0.228905* (0.0474) | -0.1495* (0.0598) |
| Income | 0.011872* (0.0447) | 0.009439* (0.0573) | -0.458641 (0.3425) | 0.296641 (0.1530) | 0.267590 (0.3456) | 0.225881 (0.6197) | 0.111417 (0.7636) | -0.18673 (0.3569) |
| Adjusted $R^2$ | 0.76 | 0.72 | 0.68 | 0.73 | 0.72 | 0.69 | 0.71 | 0.70 |
| *Event window [+1; +10]* | | | | | | | | |
| Constant | -0.69819 (0.6079) | -0.76422* (0.0111) | -0.17454* (0.0597) | -0.52364 (0.1621) | -0.5087* (0.0719) | 0.72374 (0.2369) | -0.34853* (0.0145) | -0.510** (0.0052) |
| Trump | 0.136414* (0.0425) | 0.03571** (0.0058) | 0.132134* (0.0326) | 0.06822* (0.0519) | 0.1143** (0.0039) | 0.15123** (0.0032) | -0.432937* (0.0757) | -0.3186* (0.0436) |
| Size | -0.11819* (0.0556) | -0.0808** (0.0086) | -0.01213* (0.0538) | -0.092** (0.0091) | -0.1256* (0.0987) | 0.486343 (0.4007) | -0.15750* (0.0833) | -0.0819* (0.0475) |
| Income | 0.477612 (0.3151) | 0.30338 (0.6116) | 0.523564 (0.2200) | -0.38591 (0.3690) | 0.247740 (0.9263) | 0.224828 (0.7170) | -0.112774 (0.3336) | -0.28015 (0.3597) |
| Adjusted $R^2$ | 0.69 | 0.73 | 0.75 | 0.66 | 0.71 | 0.68 | 0.72 | 0.77 |
| **S&P 500** | | | | | | | | |
| *Event day [0 ; 0]* | | | | | | | | |
| Constant | 0.123325 (0.7852) | 0.022084 (0.9230) | -0.158571 (0.8568) | 0.681004 (0.9246) | 1.902*** (0.0001) | -0.14387* (0.0627) | 0.768447 (0.5001) | 1.324402 (0.3210) |
| Trump | -0.22762* (0.0140) | -0.17314** (0.0060) | 0.06725** (0.0015) | -0.107998 (0.9217) | -0.04*** (0.0003) | -0.09194* (0.0453) | -0.164791* (0.0577) | -0.2198* (0.0705) |
| Size | -0.091*** (0.0002) | -0.032545* (0.0858) | -0.018*** (0.0007) | 0.467872 (0.8537) | -0.063** (0.0067) | 0.077173 (0.8202) | -0.02335* (0.0140) | 0.501412 (0.1683) |
| Income | 0.8652 (0.5432) | 0.046024 (0.8430) | 0.0101*** (0.0000) | 0.467872 (0.7703) | 1.37319 (0.6280) | 0.072543 (0.8246) | 0.574093 (0.6633) | 0.49106 (0.6697) |
| Adjusted $R^2$ | 0.80 | 0.77 | 0.76 | 0.74 | 0.78 | 0.81 | 0.79 | 0.76 |
| *Event window [+1; +10]* | | | | | | | | |
| Constant | -0.410881 (0.2782) | 1.719321** (0.0053) | -1.1352** (0.0025) | 0.830045 (0.2482) | -0.57009 (0.1130) | 0.543286 (0.5308) | 4.9476*** (0.0000) | 0.28161 (0.2524) |
| Trump | 0.154489* (0.0696) | 0.069904** (0.0074) | 0.1213* (0.0398) | 0.08754** (0.0015) | 0.10693* (0.0580) | 0.202787* (0.0826) | -0.1938*** (0.0003) | -0.2510* (0.0975) |
| Size | -0.04763* (0.0364) | 0.778487 (0.4319) | -0.0415* (0.0749) | 0.110998 (0.8754) | -0.52455 (0.2938) | -0.07939* (0.0910) | -0.04804* (0.0156) | -0.082** (0.0063) |
| Income | -0.453015 (0.1288) | 0.009104* (0.0355) | 0.0193* (0.0670) | 0.159883 (0.8280) | -0.50311 (0.3755) | -0.25518 (0.6731) | 0.6702 (0.3561) | 0.250096 (0.3995) |
| Adjusted $R^2$ | 0.79 | 0.78 | 0.77 | 0.74 | 0.80 | 0.76 | 0.75 | 0.71 |

| | Nasdaq Composite | | | | | | | |
|---|---|---|---|---|---|---|---|---|
| | Event day [0 ; 0] | | | | | | | |
| *Constant* | 0.347377 | -0.017100 | 0.565629 | 0.891338* | 0.626499 | 0.45864** | 1.861390 | 0.250209 |
| | (0.2900) | (0.9819) | (0.9331) | (0.0315) | (0.1330) | (0.0025 | (0.1835) | (0.5616) |
| *Trump* | -0.28365* | -0.147745* | 0.044381* | -0.1456** | -0.0509* | -0.0828** | -0.29489** | -0.310** |
| | (0.0991) | (0.0806) | (0.0210) | (0.0027) | (0.0465) | (0.0014) | (0.0016) | (0.0042) |
| *Size* | -0.03568* | -0.044272* | -0.0889** | 0.454829 | 0.186233 | 0.142460 | -0.099722* | -0.0290* |
| | (0.0799) | (0.0929) | (0.0083) | (0.2674) | (0.6608) | (0.8497) | (0.0343) | (0.0474) |
| *Income* | 0.531872 | 0.033039* | 0.896641 | 0.225881 | 0.309439 | 0.267590 | 0.023482** | 0.111417 |
| | (0.2447) | (0.0309) | (0.3530) | (0.6197) | (0.4634) | (0.7445) | (0.0095) | (0.7636) |
| Adjusted $R^2$ | 0.79 | 0.77 | 0.81 | 0.86 | 0.82 | 0.78 | 0.77 | 0.80 |
| | Event window [+1; +10] | | | | | | | |
| *Constant* | -0.098197 | -0.508746* | -8.841*** | -0.5052** | -0.7664* | 0.723704 | -2.409586 | -0.3179* |
| | (0.6079) | (0.0719 | (0.0004) | (0.0052) | (0.0111) | (0.2369 | (0.1673) | (0.0224) |
| *Trump* | 0.136414* | 0.044386** | 0.03986** | 0.10863* | 0.1135** | 0.192123* | -0.377970* | -0.3485* |
| | (0.0425) | (0.0039) | (0.0089) | (0.0436) | (0.0058) | (0.0632) | (0.0292) | (0.0145) |
| *Size* | -0.01819* | 0.125603 | -0.075701 | -0.01093* | -0.090** | -0.08634* | -0.043457* | -0.0329* |
| | (0.0556) | (0.7987 | (0.3316) | (0.0475) | (0.0086) | (0.0307) | (0.0475) | (0.0757) |
| *Income* | 0.00761* | 0.047740 | 0.141960 | 0.08015 | -0.9033 | 0.224828 | 0.664509 | -0.15755 |
| | (0.0151) | (0.9263 | (0.2725) | (0.2697) | (0.2116) | (0.7170 | (0.8312) | (0.1833) |
| Adjusted $R^2$ | 0.76 | 0.75 | 0.75 | 0.72 | 0.70 | 0.81 | 0.83 | 0.78 |

Notes: All regressions are controlled for heteroskedasticity and the p-values are given in parentheses.
∗, ∗∗, ∗∗∗ denote statistical significance at the 10%, 5% and 1% levels, respectively.

# Table 2. Sectoral impacts of 2016 U.S. presidential election on U.S. stock markets: Inclusion of further control variables

| | Financials | Oil and gas | Real estate | Consumer goods & services | Defense | Health care | Technology | Utilities |
|---|---|---|---|---|---|---|---|---|
| **Dow Jones Industrial Average** | | | | | | | | |
| **Event day [0 ; 0]** | | | | | | | | |
| Constant | 1.6223*** | -1.613535* | -1.0468* | -0.5408* | 0.9785** | 0.663966 | 0.796386* | 1.108502 |
| | (0.0000) | (0.0164) | (0.0154) | (0.0439) | (0.0041) | (0.1700) | (0.0304) | (0.1989) |
| Trump | -0.162*** | -0.171225* | 0.063*** | -0.10543 | -0.048** | 0.07307* | -0.18619** | -0.2359* |
| | (0.0000) | (0.0200) | (0.0007) | (0.7025) | (0.0037) | (0.0174) | (0.0074) | (0.0474) |
| Size | -0.085*** | 0.100477 | -0.046*** | -0.27732 | -0.07*** | -0.0530** | -0.08249** | -0.0919* |
| | (0.0000) | (0.5297) | (0.0003) | (0.5309) | (0.0008) | (0.0040) | (0.0012) | (0.0377) |
| Income | 0.0157*** | 0.478110 | -1.3087 | -0.812805 | 0.510256 | 0.006439* | 0.002378** | 0.10406 |
| | (0.0000) | (0.2239) | (0.5076) | (0.7197) | (0.2346) | (0.0233) | (0.0028) | (0.5851) |
| VIX | -0.226*** | -0.122072 | -0.111313 | -0.09861* | -0.1158* | -0.09687* | -0.14102** | -0.0984* |
| | (0.0000) | (0.1927) | (0.2324) | (0.0865) | (0.0137) | (0.0672) | (0.0059) | (0.0126) |
| Gold | -0.138*** | -0.134937* | -0.081748 | -0.05799* | -0.1217* | -0.1078* | -0.09369** | -0.0931* |
| | (0.0000) | (0.0550) | (0.4473) | (0.0153) | (0.0173) | (0.0306) | (0.0016) | (0.0460) |
| Silver | -0.026*** | -0.049471* | -0.236187 | -0.0246** | -0.0221* | 0.092213 | -0.00369** | 0.016201 |
| | (0.0000) | (0.0279) | (0.2954) | (0.0050) | (0.0235) | (0.2164) | (0.0036) | (0.2761) |
| Bitcoin | -0.249*** | -0.153349* | -0.18407* | -0.141*** | -0.129** | 0.543518 | -0.144249* | -0.1102* |
| | (0.0002) | (0.0591) | (0.0885) | (0.0007) | (0.0095) | (0.4610) | (0.0131) | (0.0202) |
| Adjusted $R^2$ | 0.93 | 0.92 | 0.88 | 0.90 | 0.87 | 0.86 | 0.91 | 0.92 |
| **Event window [+1; +10]** | | | | | | | | |
| Constant | 0.526160 | 1.162812 | 1.1717*** | -0.27057* | -1.059** | -0.32362 | 0.255759* | 0.235810 |
| | (0.4906) | (0.5185) | (0.0000) | (0.0426) | (0.0071) | (0.7345) | (0.0855) | (0.6221) |
| Trump | 0.160213* | 0.08235** | 0.1442*** | 0.08889* | 0.1239** | 0.1755** | -0.35759* | -0.262** |
| | (0.0941) | (0.0080) | (0.0000) | (0.0261) | (0.0010) | (0.0038) | (0.0705) | (0.0059) |
| Size | 0.028896* | 0.157778 | 0.0218*** | 0.0032** | 0.0104** | 0.30766 | -0.498615 | 0.0064* |
| | (0.0137) | (0.5319) | (0.0000) | (0.0030) | (0.0022) | (0.3396) | (0.8096) | (0.0152) |
| Income | 0.00154* | 0.149428 | 0.0068*** | 0.4612 | -0.879 | -0.06164 | -0.005152 | 0.6190 |
| | (0.0257) | (0.5187) | (0.0000) | (0.2353) | (0.1168) | (0.2067) | (0.9976) | (0.3028) |
| VIX | -0.0973** | -0.12142** | -0.110*** | -3.3430 | -0.0588* | -0.1569* | -0.46367** | 0.3756* |
| | (0.0052) | (0.0038) | (0.0000) | (0.2693) | (0.0316) | (0.0341) | (0.0032) | (0.0168) |
| Gold | -0.0688** | 0.301423 | -0.111*** | -0.1017* | -0.086** | -0.0667** | -0.09479* | -0.0764* |
| | (0.0015) | (0.2356) | (0.0000) | (0.0933) | (0.0046) | (0.0087) | (0.0943) | (0.0230) |
| Silver | -0.0107** | -0.00310* | -0.001*** | -0.00361* | -0.59934 | -0.0060** | -0.00266* | -0.0068* |
| | (0.0011) | (0.0372) | (0.0000) | (0.0302) | (0.3012) | (0.0079) | (0.0780) | (0.0943) |
| Bitcoin | -0.1782** | -0.092282* | -0.131*** | -0.1093** | -0.1049* | -0.152*** | -0.98847 | -0.14*** |
| | (0.0012) | (0.0441) | (0.0000) | (0.0087) | (0.0743) | (0.0005) | (0.1298) | (0.0002) |
| Adjusted $R^2$ | 0.88 | 0.86 | 0.89 | 0.90 | 0.87 | 0.88 | 0.91 | 0.86 |
| **S&P 500** | | | | | | | | |
| **Event day [0 ; 0]** | | | | | | | | |
| Constant | 0.580116 | 0.748055 | 0.402721 | 0.338153 | -1.18*** | -0.6286* | 0.565019 | 0.847395 |
| | (0.5071) | (0.3617) | (0.7487) | (0.3371) | (0.0000) | (0.0109) | (0.2963) | (0.6548) |
| Trump | -0.15359* | -0.1713** | 0.04688* | -0.1181** | -0.03*** | 0.05231* | -0.2684** | -0.3048* |
| | (0.0739) | (0.0080) | (0.0327) | (0.0026) | (0.0000) | (0.0218) | (0.0077) | (0.0465) |
| Size | 0.308786 | 1.180459 | 0.146793 | -0.02980* | -0.006** | -0.11723 | 0.069456 | -0.0090* |
|  | (0.7400) | (0.1588) | (0.1538) | (0.0207) | (0.0014) | (0.4120) | (0.2391) | (0.0101) |
| Income | 0.173006 | 1.116097 | 1.618131 | 0.002198* | 0.008*** | -0.180776 | 0.259222 | 0.0094* |
|  | (0.8519) | (0.1910) | (0.2087) | (0.0185) | (0.0000) | (0.4638) | (0.6067) | (0.0899) |
| VIX | -0.14850* | -0.12156* | -0.12728* | -0.06072* | -0.09*** | -0.05521* | -0.15945** | -0.3990* |
|  | (0.0706) | (0.0317) | (0.0245) | (0.0105) | (0.0000) | (0.0955) | (0.0079) | (0.0897) |
| Gold | -0.09872* | -0.117354* | -0.1404** | -0.072*** | -0.13*** | -0.10638* | -0.14314** | -0.0258* |
|  | (0.0603) | (0.0155) | (0.0060) | (0.0091) | (0.0000) | (0.0140) | (0.0039) | (0.0848) |
| Silver | -0.00837* | -0.00389* | -0.0051** | -0.004*** | -0.001** | -0.00842* | -0.53942 | -0.0046* |
|  | (0.0825) | (0.0250) | (0.0030) | (0.0002) | (0.0012) | (0.0714) | (0.3617) | (0.0781) |
| Bitcoin | -0.1706** | -0.21649** | -0.187*** | -0.192*** | -0.103** | -0.0906** | -0.10261* | -0.1475* |
|  | (0.0047) | (0.0091) | (0.0002) | (0.0001) | (0.0046) | (0.0019) | (0.0963) | (0.0498) |
| Adjusted $R^2$ | 0.88 | 0.89 | 0.90 | 0.91 | 0.93 | 0.87 | 0.89 | 0.94 |
| Event window [+1; +10] |||||||||
| Constant | -1.636*** | -1.96539 | -0.40923* | 0.521058 | -0.4009* | -1.654*** | 1.201386** | 0.347377 |
|  | (0.0000) | (0.0000) | (0.0352) | (0.3894) | (0.0305) | (0.0000) | (0.0037) | (0.2900) |
| Trump | 0.126*** | 0.06539*** | 0.11132* | 0.063870 | 0.1027** | 0.1665*** | -0.31386** | -0.3836* |
|  | (0.0000) | (0.0000) | (0.0294) | (0.1884) | (0.0066) | (0.0000) | (0.0064) | (0.0991) |
| Size | -0.009*** | -8.986810 | -0.046*** | 0.634510 | -0.0016* | -0.0013** | -0.0041*** | 0.635568 |
|  | (0.0000) | (0.0000) | (0.0002) | (0.1843) | (0.0133) | (0.0045) | (0.0001) | (0.1799) |
| Income | 0.0130*** | -0.98681 | 0.0024*** | 0.290433 | -0.10729 | -0.488765 | 0.00416*** | 0.131872 |
|  | (0.0000) | (0.0000) | (0.0000) | (0.5893) | (0.3732) | (0.6532) | (0.0000) | (0.2447) |
| VIX | -0.144*** | -0.1432*** | -0.130*** | -0.16667* | -0.1478* | -0.119*** | 0.09653*** | -0.1264* |
|  | (0.0000) | (0.0000) | (0.0000) | (0.0290) | (0.0997) | (0.0000) | (0.0001) | (0.0330) |
| Gold | -0.090*** | -0.0533*** | -0.110*** | -0.09952* | -0.1392* | -0.137*** | -0.0695*** | -0.059** |
|  | (0.0000) | (0.0000) | (0.0000) | (0.0598) | (0.0851) | (0.0000) | (0.0008) | (0.0065) |
| Silver | -0.002*** | -0.0042*** | -0.0031** | -0.0035** | -0.0193* | -0.078*** | -0.0194*** | -0.018** |
|  | (0.0000) | (0.0000) | (0.0010) | (0.0099) | (0.0166) | (0.0000) | (0.0002) | (0.0068) |
| Bitcoin | -0.105*** | -0.1047*** | -0.148*** | -0.10351* | -0.0939* | -0.086*** | -0.09140** | -0.1094* |
|  | (0.0000) | (0.0000) | (0.0000) | (0.0950) | (0.0385) | (0.0000) | (0.0095) | (0.0634) |
| Adjusted $R^2$ | 0.91 | 0.88 | 0.86 | 0.94 | 0.91 | 0.89 | 0.90 | 0.90 |
| **Nasdaq Composite** |||||||||
| Event day [0 ; 0] |||||||||
| Constant | -1.01718 | 0.565629 | 0.891338* | -0.17745* | -0.09819 | -0.50874* | -0.8418*** | -0.505** |
|  | (0.9819) | (0.9331) | (0.0315) | (0.0597) | (0.6079) | (0.0719 | (0.0004) | (0.0052) |
| Trump | -0.17745* | -0.204381 | 0.045861 | -0.11213* | -0.0164* | 0.084386 | -0.23986** | -0.3186* |
|  | (0.0806) | (0.9210) | (0.1527) | (0.0326) | (0.0425) | (0.0039) | (0.0089) | (0.0436) |
| Size | -0.01427* | -0.088889 | 0.054829 | -0.01213* | -0.1181 | 0.125603 | -0.175701 | -0.1109 |
|  | (0.0929) | (0.8403) | (0.2674) | (0.0538) | (0.2556) | (0.7987 | (0.3316) | (0.4475) |
| Income | 0.003039* | 0.006641 | 0.025881 | -0.023564 | 0.0076* | 0.047740 | -0.141960 | 0.0001* |
|  | (0.0309) | (0.3530) | (0.6197) | (0.2200) | (0.0151) | (0.9263) | (0.2725) | (0.0697) |
| VIX | -0.0864** | -0.10139* | -0.0529** | -0.1235** | -0.1164* | 0.723704 | -0.10958* | -0.1179* |
|  | (0.0025) | (0.0835) | (0.0016) | (0.0021) | (0.0111) | (0.2369) | (0.0673) | (0.0224) |
| Gold | -0.0828** | -0.154893* | 0.650977 | -0.1682** | -0.135** | 0.792123 | -0.07970* | -0.1285* |
|  | (0.0014) | (0.0216) | (0.1142) | (0.0079) | (0.0058) | (0.1632) | (0.0292) | (0.0145) |
| Silver | 0.142460 | -0.009722* | 0.228905 | -0.492015 | -0.0063* | -0.0063** | -0.003457* | -0.0129* |
|  | (0.8497) | (0.0343) | (0.5474) | (0.2691) | (0.0486) | (0.0054) | (0.0475) | (0.0757) |
| Bitcoin | -0.13759* | -0.14348** | 0.15141* | -0.0859** | -0.1433* | -0.12482* | -0.14509* | -0.1575 |
|  | (0.0445) | (0.0095) | (0.0636) | (0.0090) | (0.0116) | (0.0170) | (0.0312) | (0.1833) |
| Adjusted $R^2$ | 0.89 | 0.88 | 0.90 | 0.84 | 0.89 | 0.90 | 0.89 | 0.87 |

| | | | | Event window [+1; +10] | | | | | |
|---|---|---|---|---|---|---|---|---|---|
| Constant | 0.3669*** | 0.2875* | 0.561309 | 1.01605 | 0.6805 | 0.20184** | 0.11768*** | 0.2413** |
| | (0.0000) | (0.0891) | (0.3556) | (0.4934) | (0.8022) | (0.0032) | (0.0006) | (0.0022) |
| Trump | 0.1669*** | 0.039852** | 0.156533* | 0.07958** | 0.1328** | 0.18036* | -0.34765* | -0.3615* |
| | (0.0000) | (0.0097) | (0.0939) | (0.0040) | (0.0099) | (0.0124) | (0.0110) | (0.0421) |
| Size | -0.0109** | -0.028864* | 0.365453 | 6.587480 | 0.152158 | 0.003 | 0.236 | -0.056* |
| | (0.0038) | (0.0433) | (0.6230) | (0.1445) | (0.6128) | (0.7651) | (0.5592) | (0.0133) |
| Income | 0.00199** | 0.626058 | 0.153943 | 0.006572* | 0.006276 | 0.3176 | 0.141541 | 0.00213* |
| | (0.0026) | (0.1017) | (0.8581) | (0.0386) | (0.9833) | (0.1056) | (0.5518) | (0.0625) |
| VIX | -0.1157** | 0.584548 | -0.08684* | -0.1467** | 0.295877 | 0.11523* | -0.0806** | -0.076** |
| | (0.0028) | (0.1119) | (0.0227) | (0.0047) | (0.3136) | (0.0904) | (0.0089) | (0.0018) |
| Gold | -0.1295** | -0.07967** | 0.327995 | -0.10747* | -0.1355* | -0.1048** | -0.09643** | -0.042** |
| | (0.0026) | (0.0011) | (0.6996) | (0.0162) | (0.0763) | (0.0061) | (0.0057) | (0.0049) |
| Silver | -0.0085** | -0.00166** | -0.0054** | -0.00781* | 0.184704 | -0.00345* | -0.0032*** | -0.01*** |
| | (0.0049) | (0.0020) | (0.0000) | (0.0389) | (0.5558) | (0.0512) | (0.0005) | (0.0002) |
| Bitcoin | -0.1486* | -0.14459* | -0.1730** | -0.10645* | -0.1319* | -0.1249* | -0.1042** | -0.1245* |
| | (0.0141) | (0.0301) | (0.0035) | (0.0179) | (0.0317) | (0.0617) | (0.0058) | (0.0950) |
| Adjusted $R^2$ | 0.92 | 0.90 | 0.89 | 0.85 | 0.83 | 0.87 | 0.91 | 0.88 |

Notes: All regressions are controlled for heteroskedasticity and the p-values are given in parentheses.
∗, ∗∗, ∗∗∗ denote statistical significance at the 10%, 5% and 1% levels, respectively.